\documentclass[a4paper,12pt]{article}

\usepackage{amssymb,amsfonts,amsmath}

\title{An indicator energy  of two close  levels}
\author{Alexander V. Shamanin\footnote{E-mail: f2xm1@ya.ru, Russia.}}

\begin{document}
\maketitle
\thispagestyle{empty} 

\begin{abstract}

In this paper, we introduce a concept of an indicator energy of two close levels in the perturbation. \\
\\
Keywords: Quantum mechanics, perturbation theory, two close levels

\end{abstract}

\vspace{1cm}

\section{Introduction}
This paper introduces the concept of the indicator energy of two close levels in the (time-independent) perturbation. In its content, this material does not contain any complicated calculations and has a small methodological side in own nature.

Unlike a conventional approach is that we do not focus on the removal of the degeneracy by the perturbation, but on the reverse that process.

The introduction of the indicator  energy helps to classify the different cases of the behavior of levels under the influence of the perturbation, and also allows us to consider various perturbations from a single viewpoint. In addition, this concept can be used to analyze the shape of the perturbations.

It should be noted also that the section ``Formal justification'' was introduced with illustrative purpose.

\section{Formal justification}
\sloppy

Let the system with a Hamiltonian $ \hat{H}_0 $ is degenerate. We choose a time-independent small perturbation  $ \hat{V} $, that in a system with a Hamiltonian  $ \hat{H} = \hat{H}_0 + \hat{V} $ the degeneracy is lifted.

We now choose for a system with the Hamiltonian $ \hat{H'_0} = \hat{H}_0 + \hat{V} $ the perturbation in the form $ \hat{V'} \! = \! -k \hat{V,} \quad k \in [0,1]  $.
Thus, we get a system with the Hamiltonian $ \hat{H'} = \hat{H'_0} - k \hat{V} $.

It is clear that applying to the non-degenerate system with the Hamiltonian $ \hat{H'_0} $ the perturbation $ \hat{V'} $ for $ k=1 $, we obtain the system with the Hamiltonian $ \hat{H}_0 $, that is, the system with a degeneration.
(Intermediate values of $ k $ introduced as an illustration of the possibility of a continuous transition from the non-degenerate system to the degenerate system.)

In fact, the above situation is artificial. However, it has the direct bearing to the mathematical formalism of perturbation theory and it should correctly be taken into account.

\section{Can some perturbation not   lead to the repulsion of the levels?}
\sloppy

It has been possible, without relying explicitly on the above artificial situation, find a form of the perturbation 
that two close levels of the system converge.

Considered states  $ \vert l \rangle , \vert l+1 \rangle $ are non-degenerate, belong to a discrete spectrum and normalised to unity $ \langle m \vert n \rangle = \delta_{mn} $.
A simplified notation within the paper:
\begin{equation*}
\begin{array}{lll}
E^{(0)}_l \equiv E^0_1 , \: \vert l \rangle \equiv \vert 1 \rangle ,   \\ 
E^{(0)}_{l+1} \equiv E^0_2 , \, \vert {l+1} \rangle \equiv \vert 2 \rangle . 
\end{array}
\end{equation*}

Saying  ``two close levels'', we mean that the unperturbed difference $ \Delta E^0 = E^0_2 - E^0_1 > 0 $ is much smaller than the distances to other levels of the system.

The Hamiltonian of the perturbed system
\begin{equation*}
\hat{H} = \hat{H}_0 + \hat{V,} \quad \hat{H} \vert \psi_{\scriptscriptstyle{E}} \rangle = E \vert \psi_{\scriptscriptstyle{E}} \rangle  \quad \;   \\
\end{equation*}
and, 
\begin{equation*}
\begin{array}{lll}
H_{11} = \langle 1 \vert \hat{H}_0 + \hat{V} \vert 1 \rangle = E^0_1 + V_{11},   \\
H_{12} = \langle 1 \vert \hat{H}_0 + \hat{V} \vert 2 \rangle = V_{12} \quad \text{and etc.}
\end{array} 
\end{equation*}

As the state vectors  $ \vert \psi^{(0)}_1 \rangle $ and $ \vert \psi^{(0)}_2 \rangle $ are the zero approximation, we take no vectors $ \vert 1 \rangle $ and $ \vert 2 \rangle $, and their linear combinations:
\begin{equation}
\left\{ \begin{array}{lcc}
\vert \psi^{(0)}_1 \rangle = a \vert 1 \rangle + b\vert 2 \rangle  \\
\vert \psi^{(0)}_2 \rangle = c \vert 1 \rangle + d\vert 2 \rangle  ,
\end{array} \right. 
\end{equation}
where $ a,b,c,d \in \mathbb{C}, $
and the perturbed Schrodinger equation for the considered close  levels, we write in the form
\begin{equation}
\hat{H} \vert \psi^{(0)}_k \rangle = E_k \vert \psi^{(0)}_k \rangle , \; k=1,2;
\end{equation}
here by $ E_{1,2} $ denote the energies of  perturbed levels in the  corresponding (1) approximation.

It fallows from (1) and (2) that
\begin{equation}
\left\{ \begin{array}{lcc}
(H_{11} - E_1) a + H_{12} b = 0   \\
H_{21} a + (H_{22} - E_1)b = 0 \,  
\end{array} \right. 
\end{equation}
and
\begin{equation}
\left\{ \begin{array}{lcc}
(H_{11} - E_2) c + H_{12} d = 0   \\
H_{21} c + (H_{22} - E_2)d = 0 ,   
\end{array} \right. 
\end{equation}
It is seen that system (3) becomes (4), if we replace
\begin{equation*}
\begin{array}{ccc} 
E_1 \longrightarrow E_2, \\
a \longrightarrow c,  \\
b \longrightarrow d,  
\end{array}
\end{equation*}
therefore, the energies $ E_1 $ and $ E_2 $ are determined from the equation
\begin{equation}
\left\vert \begin{array}{cc}
(H_{11} - E_k)  \quad H_{12}    \\
H_{21} \quad   (H_{22} - E_k)    
\end{array} \right\vert = 0, \quad k = 1,2.
\end{equation}
We can properly collate the roots of equation (5) with the required energies, if we use the condition

\begin{equation}
\lim_{\hat{V} \to 0}{E_k}  = E^0_k, \quad k =1,2.
\end{equation}
Consequently,
\begin{equation}
E_{1,2} = \frac{H_{11} + H_{22}}{2} \mp \frac{1}{2} \sqrt{(H_{22} - H_{11})^2 + 4 {\vert H_{12} \vert}^2 },
\end{equation}
where the minus sign refers to the level $ E_1 $ and the plus sign to the level $ E_2 $.

Let's see now that we get if (6) holds. The perturbed difference is:
\begin{equation}
\Delta E = E_2 - E_1 = \sqrt{(V_{22} - V_{11} + \Delta E^0)^2 + 4 {\vert V_{12} \vert}^2 }.
\end{equation}

And cases

1) \;  $ \Delta E > \Delta E^0, $  \quad ``level repulsion'' ; 

2) \;  $ \Delta E < \Delta E^0, $  \quad rapprochement of levels; 

3) \;  $ \Delta E = \Delta E^0, $  \quad distance between levels does not change; 

4) \;  $ \Delta E = 0, $  \quad \quad \; superimposition of levels; \\
are convenient to consider, if we entering on the basis of (8)  \textit{the indicator energy of two close  levels}
\begin{equation}
\varepsilon = \frac{(V_{11} - V_{22})^2 + 4 {\vert V_{12} \vert}^2}{2 (V_{11} - V_{22})}.
\end{equation}

If we put unperturbed difference $  \Delta E^0 $ and the indicator (9) on one axis, it is easily to verify that (in this order of approximation) we have the above cases:

1) if $ \varepsilon \in (- \infty, 0 ) \cup ( \Delta E^0, + \infty ), $ -- ``level repulsion'' ;

2) if $ \varepsilon \in ( 0, \Delta E^0 ), $ -- rapprochement of levels;

3) if $ \varepsilon = 0 $ or $ \varepsilon = \Delta E^0, $ -- distance between levels does not change;

4) if $ \varepsilon = \frac{\Delta E^0}{2} $ provided that $ V_{12} = 0, $  -- superimposition of levels. \\
We see that the ``level repulsion'' is the broadest class of cases, but not the only possible.

\section{Conclusion}
It should be noted that considering  concept can be generalized to a larger number of close levels as an indicator  energy of group of close levels to characterizing the concentration or the dilution of group under the influence the perturbation. Moreover, if quantum system has  two levels only, then obtained results are exact.

The principal material (on which the work  based) can be easily found in a large number of sources. References to some of these are listed below. 

The concept of indicator energy of two close levels can be used for analytical research of quantum systems.

\section{References}
1. Auletta G., Fortunato M., Parisi G.: Quantum mechanics. Cambridge, University Press, 2009.

2. Davydov A.S.: Quantum mechanics. Oxford, Rergamon Press, 1965.

3. Anton Z. Capri.: Nonrelativistic Quantum mechanics. Singapore, World Scientific, 2002.

4. Daniel R. Bes.: Quantum mechanics. Berlin, Springer, 2007.

5. Landau L.D., Lifshitz E.M.: Quantum mechanics. Oxford, Rergamon Press, 1991.
\end{document}